# Electron Modulational Instability in the Strong Turbulent Regime for an Electron Beam Propagating in Background Plasma

Haomin Sun[1,2,4*], Jian Chen[3], Igor D. Kaganovich[1], Alexander Khrabrov[1], Dmytro Sydorenko[5]

[1]Princeton Plasma Physics Laboratory, Princeton University, Princeton, New Jersey 08543, USA

[2]CAS Key Laboratory of Geospace Environment, Department of Geophysics and Planetary Science, University of Science and Technology of China, Hefei, Anhui, People's Republic of China

[3]Sino-French Institute of Nuclear Engineering and Technology, Sun Yat-sen University, Zhuhai 519082, P. R. China

[4]CAS Center for Excellence in Comparative Planetology, People's Republic of China

[5]University of Alberta, Edmonton, Alberta T6G 2E1, Canada

Corresponding Author: Haomin Sun, Email: haomins@princeton.edu

**Abstract**

We study collective processes for an electron beam propagating through a background plasma using simulations and analytical theory. A new regime where the instability of a Langmuir wave packet can grow locally much faster than ion frequency is clearly identified. The key feature of this new regime is an Electron Modulational Instability that rapidly creates a local Langmuir wave packet, which in its turn produces local charge separation and strong ion density perturbations because of the action of the ponderomotive force, such that the beam-plasma wave interaction stops being resonant. Three evolution stages of the process and observed periodic burst features are discussed. Different physical regimes in the plasma and beam parameter space are demonstrated for the first time.

It is well known that large amplitude, high frequency plasma waves are subject to strong wave-wave nonlinear interaction, such as parametric processes [1,2] and the formation of solitary structures [3,4]. The physics of nonlinear interactions involving the Langmuir waves created by an electron beam has long been a topic of great interest [5-17], with a wide range of applications in low-temperature plasma devices [18-22] and space plasmas [23-25]. The first simplified fluid model describing nonlinear Langmuir wave-wave interaction was proposed by V. E. Zakharov in 1972 [26], who predicted that the Langmuir wave packets would self-similarly focus into the smaller and smaller region when their intensity is large enough, at the same time ion density perturbations also grow due to the action of ponderomotive force. The Langmuir wave energy could be further accumulated in the density depletion regions, leading to an increase in intensity of both the Langmuir waves and ion density perturbations. This is a well-known phenomenon termed as the Langmuir collapse, which was believed to produce Strong Langmuir Turbulence (SLT) [27,28]. Starting from the original Zakharov's paper, there have been numerous follow-up publications employing the well-known Zakharov's equations to model the Langmuir collapse and the wave energy properties [29-32]. There has also been some observational evidence indicating that the Langmuir collapse plays an important role in the high-frequency wave heating in the ionosphere [24,33,34]. Despite its great success, the traditional Zakharov model could not rigorously describe the wave-wave instabilities growing much faster than the ion frequency ($\omega_{pi}$) since charge quasi-neutrality condition was assumed. In contrast, in the experimental studies where an electron beam is injected into a plasma, strongly-nonlinear wave-wave interactions could evolve much faster compared with the ion response and therefore may be independent of the ion dynamics [35]. In such a case the traditional model needs to be revised in order to describe the initial stage of the wave-wave nonlinearity of the beam generated wave packets. Another shortcoming of the Zakharov equations is that it does not self-consistently account for the plasma wave damping occurred due to transferring wave energy to superthermal electrons generated in the process [36], despite several transit-time damping models [29,37-39] have been proposed to try to mitigate this problem. The detailed study of all these effects of SLT produced by the beam necessitates kinetic simulations. Previous kinetic simulations of the Langmuir Collapse [40-42] only studied the slow (ion time scale) evolution of a wave packet set as an initial condition and the mutual interaction between the beam and wave packet was not modelled. The data resolution was also low due to the limitation of computational resources at that time. Previous experimental observations such as the nonlinear evolution of beam-plasma instability [43-45] and the beam-generated Langmuir collapse [35] could not be analyzed in sufficient details, because of the limited range of timescales and wavelengths they could measure at that time.

In this Letter, we extensively studied a new regime of Langmuir wave nonlinear interaction generated from the beam-plasma interaction for ubiquitous direct current discharges with a hot cathode using high-resolution 2D PIC simulations and analytical theory. This Letter is also a joint submission of another paper in *Physical Review E* [46], where more comprehensive descriptions of different physical regimes are provided. An electron beam is generated by thermal emission from the cathode and is accelerated by a cathode sheath [22,47,48]. Simulations results reveal that in this regime, large-amplitude localized Langmuir waves are rapidly generated via a wave-wave nonlinear process we term as Electron Modulation Instability (EMI). We observed that such an instability evolves faster than ion response and, hence, the traditional Zakharov model is not applicable. Based on this important observation, we derived new analytical relations for the threshold of the SLT for the beam-generated plasma-wave packets, which also takes into account the Landau damping and collisional effects. The obtained analytical relations are verified by comparing it with results of 57 simulation cases and, correspondingly, can be used as a scaling law predicting the onset of the SLT produced by an electron beam for future experimental and numerical studies. To the best of our knowledge, this Letter also reports the first self-consistent 2D PIC simulations of SLT in a beam-plasma system.

We model a DC discharge in slab geometry consisting of a flat cathode with thermionic emission located at $x = 0$ and an anode located at $x = L_x$ using EDIPIC-2D ([49]). Only part of cathode with width $L_y$ was modelled and the periodic boundary conditions at $y = 0$ and $y = L_y$ was used. At electrodes, fixed potentials were applied and particles are absorbed. The initial number of macro-particles for plasma electrons and ions are 800 per cell. Initial plasma density is set to $n_{p0} = n_{e0} = n_{i0} = 10^{17} m^{-3}$, and the ion temperature is $T_{i0} = 0.03 eV$ (nearly equal to the room temperature). The pressure of the background gas, argon, is $3.85 mTorr$. Here we show first two selected cases with initial bulk electron temperature $T_{e0} = 0.2eV$ for Case 1 (with EMI) and $T_{e0} = 2eV$ for Case 2 (without strong Langmuir turbulence). For both cases, an electron beam with density $n_b/n_{p0} = 0.015$ and temperature $T_{eb} = 0.2eV$ is injected from the negatively biased cathode (thermionic emission) at $t = 0ns$. The cathode is biased at $t = -80ns$ to allow the sheath to reach a steady state such that the beam energy is $E_b = 30eV$ at $t = 0ns$. The simulation domain grid $L_x \times L_y = 32mm \times 9mm$ contains $3840 cells \times 1024 cells$. Each simulation lasts for $580ns$, except Case 1 lasts for $3080ns$. The beam-neutral elastic collision frequency is $\nu_{en,elas} \approx 2.1 \times 10^7 s^{-1}$ for $E_b = 30 eV$ [50], which is small comparing with the typical growth rate of two-stream instability ( $\gamma = \sqrt{3}/2\omega_{pe}(n_b/2n_0)^{1/3} \approx 3.02 \times 10^9 s^{-1}$ ). A transformation to dimensionless units is available in supplementary material [51].

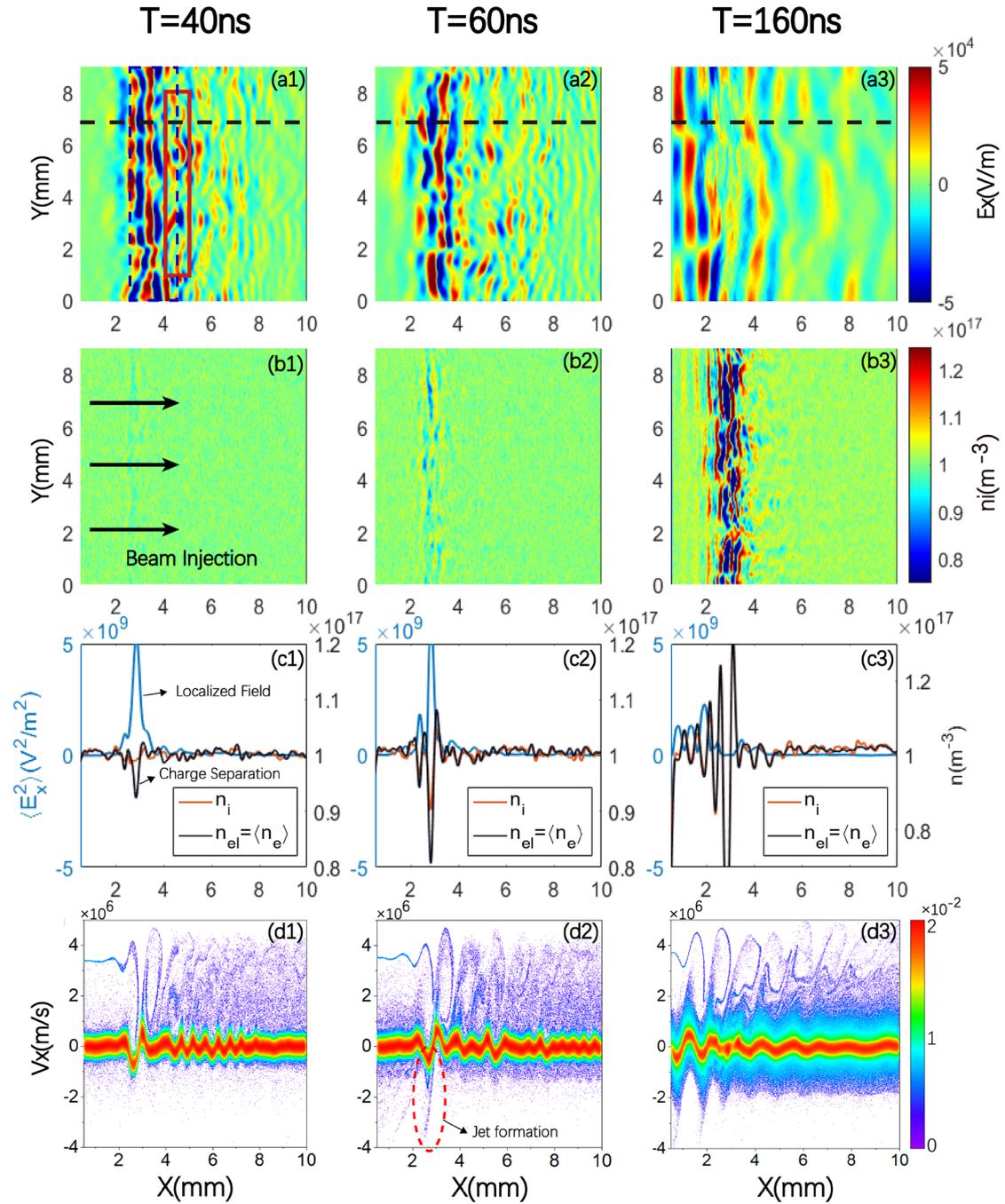

**Figure 1**: Snapshots of strong Langmuir turbulence for Case 1 (with EMI) at $t = 40ns, 60ns, 160ns$ shown for part of the simulation domain $x = (0, 10mm)$. (a1)-(a3) show the 2D color plots of the electric field $E_x$. The black dashed lines show the $y = 6.8mm$ location where we plot (c1)-(d3). The blue dashed rectangle outlines the region used to produce the line plots shown in Fig.2 (c)-(d). The red rectangle shows the region used to calculate the EVDF plotted in Fig.2 (e)-(f). (b1)-(b3) show the time evolution of the ion density profile, $n_i$. (c1)-(c3) show the $\langle E_x^2 \rangle$ and density profiles of ions and electrons along the black dashed lines, where the $\langle ... \rangle$ denotes the time average over the time interval $3.025ns$ (10 plasma periods). Note the charge separation in (c1) and (c2). (d1)-(d3) show the electron phase space along the same black dashed lines. The color bar to the right of the figure shows EVDF normalized to unity by the integration of the EVDF.

Figure 1 shows snapshots in Case 1. The electron beam is injected from the cathode [Fig.1 (b1)], interacts with the plasma and creates a large amplitude Langmuir wave packet [Fig.1 (a1) and (c1)]. At $t = 40ns$, the Langmuir wave packet has grown locally above saturation level ($|E| = 3.2 \times 10^4 V/m$ calculated by Eq. (3)), but the ion density is still almost uniform. Two characteristic features of EMI manifest here: First, the strong ponderomotive force $\nabla(\epsilon_0 E^2/4)$ of the localized field is balanced by the electrostatic force $n_{p0}eE_l$ resulting from charge separation because we can see in Fig.1 (c1)-(c2) the charge separation is existing together with the electric field envelope (for $t = 40ns$, $\epsilon_0 E^2/4n_{p0}T_e \sim (n_{el} - n_i)/(n_{p0}k^2\lambda_{De}^2) \sim 2$, where $\nabla E_l = e/\epsilon_0(n_i - n_{el})$, "$l$" denotes time average). While in the traditional Langmuir collapse, charge neutrality $\delta n_i \approx \delta n_{el}$ is assumed and the ponderomotive force $\nabla(\epsilon_0 E^2/4)$ was assumed to be balanced by thermal pressure force $\nabla(\delta n_{el} T_e)$ [45,52,53]. Second, the wave energy grows in the EMI process and forms a local peak in the smaller and smaller region before the ion moves, indicating a "localization" of Langmuir waves faster than ion frequency ($\omega_{pi}$), while the traditional Langmuir collapse process happens comparable to or slower than the ion response. Both of these two new features are beyond the applicability of the Zakharov model. The follow-up phase mixing is also evident in the phase space plot shown in Fig.1 (d1)-(d2). The maximal intensity of the Langmuir waves is almost six times of the saturation level at around $t = 60 ns$. Associated with much bigger Langmuir wave amplitudes the ion density perturbations start to grow at the locations of the electric field peaks at around $t = 46 ns$, as evident in Fig.1 (c2). At $t = 160 ns$, the intensity of the Langmuir waves has dramatically decreased, whereas the ion density perturbations have significantly grown to reach nearly 50% modulation levels [Fig.1 (c3)]. The ion density perturbation at the maximum is $\delta n_{i,max}/n_{p0} = 0.59 < \epsilon_0 |E_{peak}|^2/4 n_{p0} T_e \sim 2.5$, which further confirms that the ions don't have enough time to respond to the wave growing and the thermal pressure cannot balance the ponderomotive force. We observe electrons being accelerated in the direction opposite to the direction of beam propagation, indicating strong backward waves and wave energy trapped in the density trough; they are presented as jet formation in the electron phase space plots shown in Fig.1 (d2)-(d3).

 We see that such an intermittent behavior will finally cease with the increase of the bulk electron temperature. Note that for this case, the second burst is already in the SWMI regime, while for a narrower beam or for a larger simulation domain, the system could stay in the EMI regime for a longer time [46]. The red and yellow lines in Fig.2(b) show that the linear growth rate of two-stream instability and EMI match well with the simulation. Here, we only show the first burst period to illustrate the evolution of EMI. The evolution of the wave energy is shown in Fig.2 (c)-(d) for a comparison between Case 1, $T_{e0} = 0.2 eV$ and Case 2, $T_{e0} = 2 eV$. The nonlinear processes of wave energy evolution observed for Case 1 exhibit three stages. Stage I, $t \approx 0 - 90 ns$, is a typical period when the strong Langmuir turbulence develops during $t = 20 - 60 ns$ and decays during $t = 60 - 90 ns$. The bulk electron heating, $E \cdot J_{bulk}$, is strong in Stage I, when energy transfers from the beam to the electric field and then to the bulk electrons. The strong energy transfer was known as the "burnout" of wave packet [5,41]. Therefore, the average temperature has increased from 0.2eV to 1.07eV during $t = 0 - 90 ns$ for Case 1, whereas the temperature increased only from 2.0eV to 2.15eV for Case 2. As a result of the strong electric field, the beam scattering angle in Case 1 could reach $\theta = \arctan v_y/v_x = 30°$, marked by the white lines in Fig.2 (e) while the beam energy simply spreads along $W_x$ to the electron bulk population corresponding to the wave-particle interaction saturation [54] for Case 2. For the first time, we clearly identified a $k^{-5}$ spectrum in EMI at $t = 30ns,\ 60ns$ in Fig.3 (g) for Case 1, during which wave packet is localizing. One possible explanation is the interaction of strong turbulent Langmuir waves with the accelerated super-thermal electrons [36,55].

Because ions are heavy, it takes some time for ions to respond to the ponderomotive force. At about $t = 110ns$, the initial ion density perturbations grow to a significant value $\delta n_i/n_{i0} \sim 0.5$, when Stage II starts. During Stage II a secondary standing wave is generated at the beam injection $x < 2mm$ and the initial ion density perturbations also spread from the initial location at $x$ around $3mm$ to $x < 2mm$ in form of ion acoustic waves [see also Fig.1 (c3)] [42]. This creates a larger region with strong ion density perturbations (see movies in the supplementary material [51]). When the ion density perturbations grow to about $30\%$ near the beam injection at $x < 2mm$, the Stage III starts at $t > 260ns$. Because of the large-amplitude ion density perturbations near the beam injection point, the beam-plasma interaction stops being resonant. The plasma waves disappear in the region with strong ion density perturbations $x = 0 - 4mm$. When such ion density perturbations gradually relax, the next burst would start.

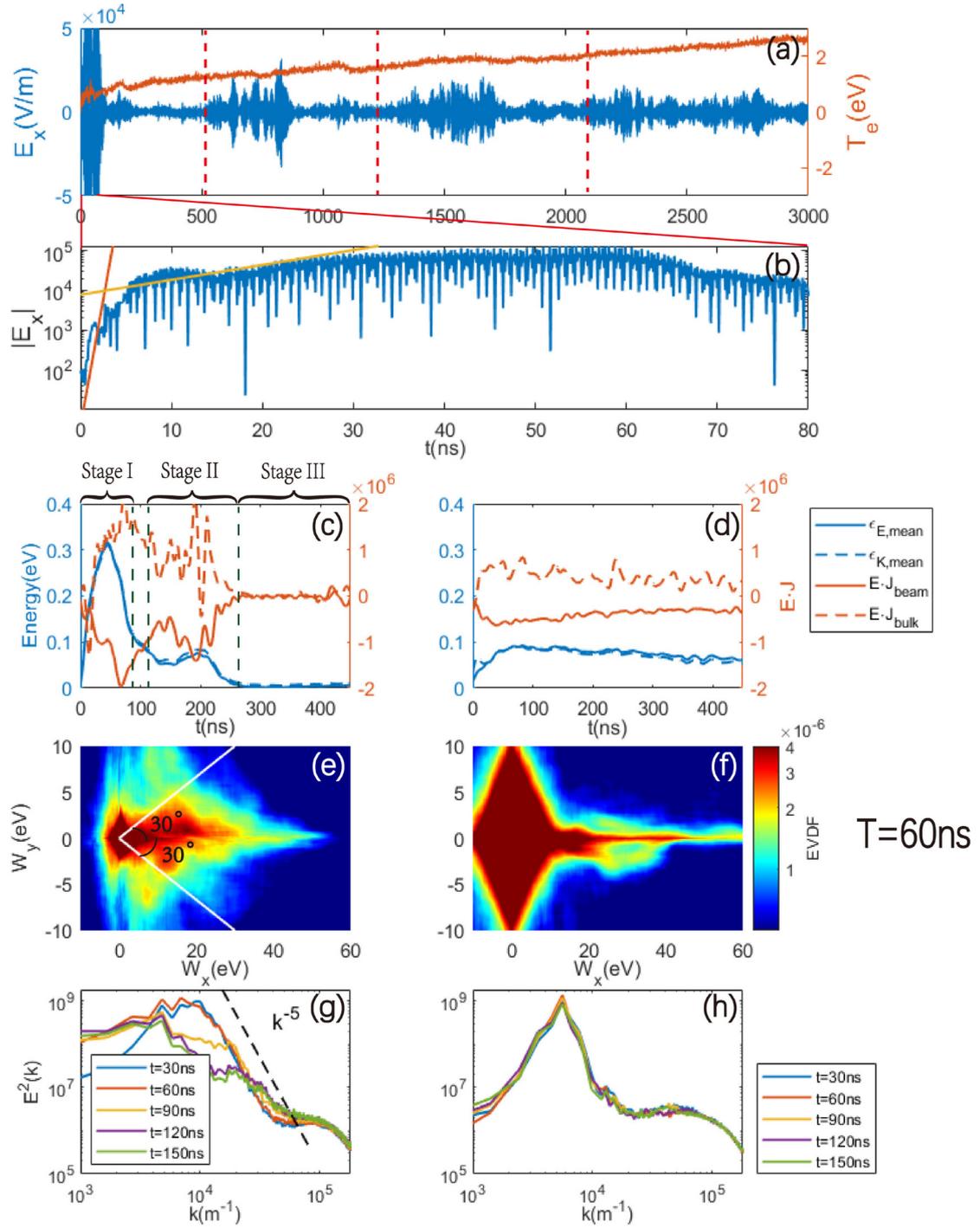

**Figure 2**: (a) shows the periodic burst feature of EMI in $E_x$ (detected by probe at $x = 2.8mm$, $y = 4.5mm$ for Case 1) and increase of average bulk electron temperature $T_e$. The three periods are roughly indicated by the red dashed lines. Note only the first burst is in the EMI regime for this case. (b) presents the time evolution of $|E_x|$ at the initial stage. The red line shows the linear growth rate of two-stream instability while the yellow line gives the EMI growth rate $\gamma_{EMI} \approx 7.4 \times$

$10^7 s^{-1} > \omega_{pi} \approx 3 \times 10^7 s^{-1}$ calculated by Eq. (19) in our accompanying paper [46]. (c) and (d) show the time evolution of the averaged electric field energy $\epsilon_{E,mean}$, averaged kinetic energy for bulk electrons $\epsilon_{K,mean}$, averaged energy transfer rate from wave to beam $E \cdot J_{beam}$ (hence negative) and averaged energy transfer rate from wave to the bulk plasma $E \cdot J_{bulk}$ during $t = 0 - 450 ns$ for Cases 1-(c) and Case 2-(d), respectively. Fig.2 (e) and (f) show colorplot of the electron velocity distribution function (EVDF) at $t = 60 ns$ for Case 1 and Case 2. Both EVDFs are normalized to unity by the integration of EVDF. (g) and (h) show temporal evolution of the energy spectrum $E^2(k)$ for Cases 1 and 2, where $k = \sqrt{k_x^2 + k_y^2}$.

From the theory perspective, the onset and initial stage of wave-wave nonlinear interaction can be approximately described by multi-fluid nonlinear wave coupling equations. Details of derivations are given in our accompanying paper [46]. The threshold of SLT onset can be obtained by balancing the ponderomotive force with pressure force:

$$\frac{\epsilon_0 |\widetilde{E}_{threshold,SWMI}|^2}{4 n_0 T_e} = max\left[(k\lambda_{De})^2, \frac{2\Gamma_e}{\omega_{pe}}\right], \qquad (1)$$

where $\Gamma_e$ is the damping rate, whose expression will be given later. This threshold differs from the well-known Zakharov threshold [26] since we also considered damping. Above this threshold, a localized standing wave begins to generate and modulate the beam-created wave packet (slower than ion frequency). We therefore call this instability Standing Wave Modulational Instability (SWMI). We also showed that there is another higher threshold for the Langmuir wave growth faster than the ion response if the electric field is so strong such that the charge neutrality condition (as is assumed in the Zakharov models) breaks down and the ponderomotive force is balanced by the electrostatic force created by charge separation (see Case 1). The threshold can be expressed by:

$$\frac{\epsilon_0 |\widetilde{E}_{threshold,EMI}|^2}{4n_0 T_e} = max\left[1 - \frac{n_b}{3n_0}\frac{1}{k^2\lambda_{De}^2}, 2\frac{\Gamma_e}{\omega_{pe}}\frac{1}{k^2\lambda_{De}^2}\right]. \tag{2}$$

Physically, it means that the electric field must be strong enough to modify the electron dynamics and create charge separation in the nonlinear process of wave concentrating into the smaller and smaller region before ions move. At the same time, the damping must be small enough so that the wave could grow locally. Since it involves only electron dynamics in the initial stage, we call it "Electron Modulational Instability" (EMI). The EMI process is essentially different from classical Langmuir collapse since it describes a faster instability. We believe it is this instability that gives the strong local Langmuir waves in Case 1.

The beam excitation of the original pump wave determines the initial saturation amplitude of the electric field $E_{sat}$ before modulational instabilities occur. In our simulations, the Quasi-Linear (QL) approach cannot describe the wave saturation and the wave-particle trapping process needs to be considered instead, see e.g., Ref. [54]. The initial saturated electric field can be estimated by:

$$\frac{\epsilon_0 E_{sat}^2}{4n_0 T_e} = \frac{9}{8}\left(\frac{n_b}{n_0}\right)^{4/3}\frac{m_e v_b^2}{2T_e}, \tag{3}$$

where $n_b$ is the beam density, and $v_b$ is the beam velocity. The saturation amplitude of the beam-generated plasma wave given by Eq. (3) has been verified experimentally [35], in other simulations [56] and our simulations (see our accompanying paper [46]).

Substituting Eq. (3) into Eq. (1) and (2) we obtain the criterion for the SLT regime:

$$\frac{9}{8}\frac{m_e v_b^2}{2T_e}\left(\frac{n_b}{n_0}\right)^{4/3} > max\left[\frac{2\Gamma_e}{\omega_{pe}}, (k\lambda_{De})^2\right]. \qquad (4)$$

And wave localization is faster than ion response:

$$\frac{9}{8}\frac{m_e v_b^2}{2T_e}\left(\frac{n_b}{n_0}\right)^{4/3} > max\left[1 - \frac{n_b}{3n_0}\frac{1}{k^2\lambda_{De}^2}, 2\frac{\Gamma_e}{\omega_{pe}}\frac{1}{k^2\lambda_{De}^2}\right]. \qquad (5)$$

To confirm predictions for the threshold (4) and (5), we further performed 57 simulations with different beam energies and initial bulk electron temperatures. As explained above, kinetic effects of the Landau damping needs to be accounted for to correctly calculate the threshold (4) and (5). Before the onset of strong turbulence, the EVDF is approximately a Maxwellian and the wave damping can be approximated by $\Gamma_e \approx \sqrt{\pi/8}\omega_{pe}/(k\lambda_{De})^3 \exp(-1.5 - 1/2/(k\lambda_{De})^2) + \nu_{en}/2$ [52], where $\nu_{en}$ is the collisional frequency between electrons and neutrals. Here, $k$ is taken to be comparable to $k_0 = \omega_{pe}/v_b$. For nonlinear evolution of SLT, several phenomenological transit-time damping models could be used in the place of linear Landau damping [29,37-39]. Figure 3 shows Eq. (4) and Eq. (5) by the blue line and red line. The blue line for Eq. (4) separates cases between the SLT (red stars) and other regimes (blue triangles). The red line for Eq. (5) separates cases with EMI (red plus-over-an-x markers) and without EMI (red stars) in a large parameter space of beam to plasma densities (two orders of magnitude).

When the damping can be neglected in Eq. (4) and (5), namely, when the beam is very energetic and the wavelength is long, the criteria can be well approximated by the following two scalings:

$$\frac{E_b}{T_e} \sim \frac{2}{3}\left(\frac{n_b}{n_0}\right)^{-\frac{2}{3}}, \quad (6)$$

$$\frac{E_b}{T_e} \sim \left(\frac{9}{8}\left(\frac{n_b}{n_0}\right)^{\frac{4}{3}} + \frac{2}{3}\frac{n_b}{n_0}\right)^{-1}. \quad (7)$$

The scaling given by Eq. (7) separates a new regime that has not been studied in detail to the best of our knowledge. The scaling given by Eq. (6) is also different from the one given by A. Galeev *et al.*, $E_b/T_e \sim (n_b/n_0)^{-1/3}$ [57], because the authors of Ref. [57] used the QL theory to estimate saturation levels of waves excited by the beam, whereas in our case the saturation mechanism is due to the wave trapping. The QL theory is valid only if $\Delta v_{bT}/v_b > (n_b/n_0)^{1/3}$, $\Delta v_{bT}$ is the beam thermal velocity spread [22], which rarely holds for most discharges with hot cathodes where $T_{eb} < 0.2 eV$ and $\Delta v_{bT}/v_b \ll (n_b/n_0)^{1/3}$ [22,35].

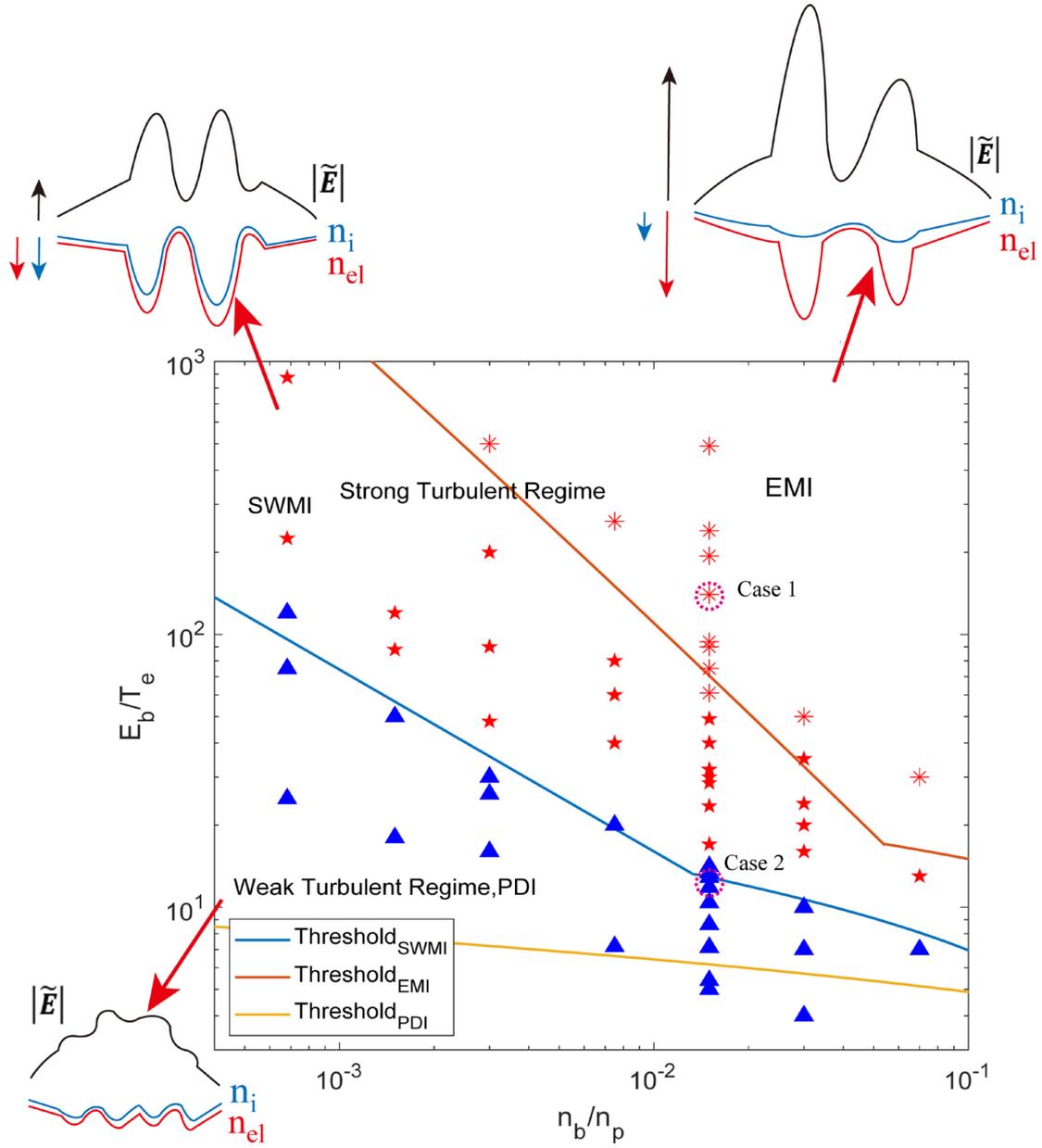

**Figure 3**: Parameter space of ratio of the beam energy to the bulk electron temperature $E_b/T_e$ versus the ratio of the beam density to the plasma density, $n_b/n_p$. The blue line shows the threshold Eq. (4) and the red line shows the threshold Eq. (5). Physical pictures of different regimes are shown ($|\tilde{E}|$ denotes wave packet). The yellow curve shows the threshold of Langmuir Parametric Decay Instability (PDI) (which comes from Ref. [1]). Red and blue markers show the cases with and without strong turbulence, respectively. Red markers are used only if the clear large amplitude

standing wave feature and the associated ion density dips are observed. Red plus-over-an-x markers denote the cases with EMI, where fast localization of Langmuir waves faster than $\omega_{pi}$ and electrostatic force resulting from charge separation that balances the ponderomotive force are clearly observed. Pink dashed circles mark Cases 1 and Case 2 used for analysis in this letter.

We identified a new regime in beam-plasma interaction process where the Electron Modulational Instability (EMI) creates a localized wave packet rapidly faster than the ion frequency as opposed to the traditional Langmuir collapse. Broad-spectrum, strong heating to bulk plasma and scattering to beam electrons in EMI regime are quantified in simulations. The SLT exhibits rapid periodic bursts ($\omega_{pe}T < 10^4$) for a system that is initially in the EMI regime. We have also proposed and verified analytical criteria (given by Eqs. (4-7)) for the onset of SLT that can explain past and guide future numerical and experimental studies of beam-plasma interactions, such as that in low-temperature ($T_e \leq 1eV$) pulsed beam systems [18-22] and certain space plasmas [24,25,58].

The authors thank referees for the careful reading and really help comments in improving this manuscript. The authors thank Prof. Ilya Dodin, Prof. Quanming Lu, Prof. Chuanbing Wang, Dr. Stephan Brunner, Dr. Justin Ball, Dr. Sarveshwar Sharma and Dr. Liang Xu for the fruitful discussions. The work of I.K., A.K, and D.S. was supported by the Princeton Collaborative Research Facility (PCRF) and Laboratory

Directed Research & Development (LDRD) projects, which are funded by the U.S. Department of Energy (DOE) under Contract No. DE-AC02-09CH11466.